\documentclass[twocolumn,prb,printnumbers,superscriptaddress,amsmath,amssymb]{revtex4}
\usepackage{graphicx}
\usepackage{color}

\begin{document}

\title{Stability dependence of local structural heterogeneities of stable amorphous solids}
\date{today}

\author{Alireza Shakerpoor}
\affiliation{Department of Chemistry, Colorado State University, Fort Collins, Colorado 80523, USA}
\author{Elijah Flenner}
\affiliation{Department of Chemistry, Colorado State University, Fort Collins, Colorado 80523, USA}
\author{Grzegorz Szamel}
\affiliation{Department of Chemistry, Colorado State University, Fort Collins, Colorado 80523, USA}

\begin{abstract}
The universal anomalous vibrational and thermal properties of amorphous solids 
are believed to be related to the local variations of the elasticity.  Recently it has been shown that the vibrational properties are
sensitive to the glass's stability. 
Here we study the stability dependence of the local elastic constants of a simulated glass former over a broad range of stabilities,
from a poorly annealed glass to a glass whose stability is comparable to laboratory exceptionally stable vapor deposited glasses. 
We show that with increasing stability the glass becomes more uniform as evidenced by a smaller variance of local elastic constants. 
We find that, according to the definition of local elastic moduli used in this work, the local elastic moduli are not spatially correlated.
\end{abstract}

\maketitle

\section{Introduction}
The vibrational modes and the low temperature thermal properties of amorphous solids are sharply different from those of 
their crystalline counterparts \cite{Zeller1971,Stephens1973,Zaitlin1975,Pohl2002}. 
The uniform structure of crystals allows for the description of the low frequency modes as if it were a classical elastic body whose properties are governed by the
elastic moduli, which forms the basis of the Debye model for the density of states. This description leads to a $T^3$ increase of the specific heat for crystalline solids due to 
the increase of the density of the vibrational modes as the square of the frequency $\omega$. Recently it was shown that the low frequency vibrational modes of
amorphous solids can be divided into a Debye term and an excess contribution that increases as the fourth power of the frequency \cite{MizunoShibaIkedaPNAS2017,Wang2019}.  The excess 
modes are spatially quasi-localized. Their spatial extent and density decrease with increasing stability. The quasi-localized character of excess
modes suggests that there might be a spatially varying local elasticity. 

Indeed, there is a large body of evidence for the existence of spatially varying local elastic constants in amorphous solids
\cite{Mizuno2013,Mizuno2016,Yoshimoto2004,Tsamados2009,Leonforte2006,Mizuno2014,Gelin2016,Caroli2019,Fan2014,MizunoMossaBarratEPL2013,Wagner2011,MizunoSilbertSperlPhysRevLett2016,Pogna2019}. 
To explain a plateau observed in the thermal conductivity around 10K for many dielectric amorphous solids, 
a Rayleigh like scattering of sound waves was assumed \cite{Zeller1971,Pohl2002}.  
This assumption posits scattering from uncorrelated defects that 
are much smaller than the wavelength of the sound wave, and these defects would naturally give rise to 
local variations of the elasticity. Further theoretical analysis assuming local variations of the elasticity 
reproduces the $\omega^4$ excess in the vibrational density of states and predicts the Rayleigh scaling $k^4$ (where $k$ is a wavevector) 
of sound attenuation \cite{Schirmacher2007,Schirmacher2010,MizunoIkedaPhysRevE2018,Wang2019}. 
The $k^4$ scaling of sound attenuation was questioned in a computer simulation study \cite{Gelin2016} and a logarithmic 
correction to the Rayleigh scaling was proposed. This correction was rationalized in terms of a power law decay of the spatial correlations
of the local elasticity. However, other simulation studies \cite{MizunoIkedaPhysRevE2018,Wang2019,Moriel2019} suggest that the logarithmic 
correction either exists only for a narrow range of wavevectors  (frequencies) or this correction is only a good description of the crossover region between the high
and low wavevector (frequency) behavior of sound attenuation. 

Pogna \textit{et al.} \cite{Pogna2019} examined sound attenuation in geologically hyperaged, ultrastable amber 
within the framework of fluctuating elasticity theory 
to establish a link between stability and the local variation of the elastic constants. They fitted the predictions of the theory
for the vibrational density of states to the experimental data and in this way obtained estimates of the relative variance of 
the local elastic constants and of a 
length scale characterizing their spatial variation. They concluded that there was a reduction
in the variation of the elastic constants by around 6\% and an increase of the characteristic length scale of around 22\% in the
hyperaged amber compared to a liquid cooled sample.
Thus, increasing stability seemingly narrows the distribution of elastic constants and increases the range of their correlations. 

However, in a very recent simulational study Caroli and Lemaitre\cite{Caroli2019} 
argued that the fluctuating elasticity theory does not describe well sound attenuation in amorphous solids. 
They based this conclusion on two results. First, they showed that the fluctuating elasticity theory predicts the $k^4$ Rayleigh scattering-like
sound damping whereas their simulations were consistent with a logarithmic correction. Second, they measured the parameters that enter into
the fluctuating elasticity theory in simulations, used them to calculate sound attenuation, and compared these predictions with
sound attenuation observed in the same simulations. They found that the predicted sound attenuation is two orders of magnitude smaller
than the observed one. The second fact implies that the fluctuating elasticity theory 
severely underestimates the magnitude of the sound attenuation even if one were to argue that the logarithmic corrections is an intermediate, finite 
wavevector feature and the sound attenuation can be described within the Rayleigh scattering picture. 

We note that it is difficult to directly probe local variations of the elasticity in experiments \cite{Wagner2011}, which 
forced Pogna \textit{et al.} to treat the relative variance of the local shear modulus as a fitting parameter.
In contrast, simulations are able to 
calculate local elastic constant using several different methods \cite{Gelin2016,MizunoIkedaPhysRevE2018,Mizuno2013}. 
Using one of these methods, 
Mizuno, Mossa, and Barrat found that the width of the distribution of local elastic constant 
correlates with sound attenuation \cite{Mizuno2014}. For their study, they continuously transformed a crystal into 
an amorphous solid by continuously changing the size ratio of a binary mixture. Using the same technique 
they also demonstrated that the thermal conductivity, the lifetime of acoustic modes, and the 
local elastic heterogeneity are correlated \cite{Mizuno2016}. This investigation, however, does not mimic the experimental 
procedure of Pogna \textit{et al.}\cite{Pogna2019} who studied the stability dependence of sound attenuation.
Importantly,  in the work of Mizuno, Mossa, and Barrat the system is changed systematically in order to 
establish the correlations between the transport and acoustic properties and the variation of local elastic constants. 

Here we examine the dependence of local elastic moduli of a simulated polydisperse glass former on its stability.  
We partition the system into different box sizes $w$ and determine the distribution 
of local elastic moduli for three values of $w$. We find that the width of the distribution decreases with 
increasing stability. However, using our definition of the local elastic moduli, we find that the local elastic moduli are uncorrelated in space.  

\section{Methods}
\subsection{Molecular Dynamics Simulations}

We studied a system of $N = 48000$ and $N=192000$ polydisperse repulsive particles 
in a cubic box of volume $V$ with periodic boundaries in 3D. 
The pair potential is given by
\begin{equation}
U(r_{ij})=
\begin{cases}
\epsilon\left(\frac{\sigma_{ij}}{r_{ij}}\right)^{12} + v(r_{ij}),\hspace{20pt}&\frac{\sigma_{ij}}{r_{ij}}<r_{\text{cut}}\\[10pt]
0,\hspace{20pt}&\frac{\sigma_{ij}}{r_{ij}}\geq r_{\text{cut}}
\end{cases}
\end{equation}
with
\begin{equation}
v(r_{ij}) = c_0 + c_2\left(\frac{r_{ij}}{\sigma_{ij}}\right)^{2} + c_4\left(\frac{r_{ij}}{\sigma_{ij}}\right)^{4}.
\end{equation}
The distance between particle $i$ and particle $j$ is $r_{ij}=|$\textbf{r}$_i-$\textbf{r}$_j|$, 
$\sigma_{ij}=\frac{\sigma_{i}+\sigma_{j}}{2}\left(1-0.2|\sigma_{i}-\sigma_{j}|\right)$\cite{Wang2019,Ninarello2017}. 
The size of an individual particles $\sigma$ are given by the probability distribution
\begin{equation}
P(\sigma)=\frac{A}{\sigma^{3}}
\end{equation}
where $\sigma\in [0.73, 1.63]$ and zero otherwise. 
The coefficients $c_0$, $c_2$, and $c_4$ are chosen to guarantee the continuity of the potential up to the second derivative at the cutoff distance $r_{\text{cut}}=1.25$. 
This choice of system and polydispersity inhibits crystallization and fractionation while allowing the swap Monte Carlo algorithm to 
equilibrate to low temperatures \cite{Ninarello2017}. We present the results in reduced units with $\epsilon$ being our unit of energy, the average 
of $\sigma = \sigma_0$ being our unit of length, and $\sqrt{m \sigma_0^2/\epsilon}$ being the unit of time. 

For each  parent temperature $T_{\text{p}}\in [0.062, 0.200]$ we studied 4 independent initial configurations at number density $\rho=1$. 
Each configuration was first 
equilibrated at its parent temperature and then quenched to an inherent structure via the conjugate gradient algorithm. 
For reference, for our system the mode-coupling temperature $T_{\text{MCT}}\approx 0.108$ and the glass transition temperature $T_g\approx 0.072$ \cite{Ninarello2017}.
The equilibration was done using the swap Monte Carlo algorithm that combines conventional Monte Carlo moves with particle swaps 
\cite{Grigera2001,Gutierrez2015,Ninarello2017}. 

After quenching, we ran very low temperature NVT molecular dynamics simulations using LAMMPS \cite{lammpsURL,lammpsPUB} code to which we 
added the interaction potential for the present model. 
The time step for all of MD simulations was $dt=0.02$. We first 
ran short equilibration runs at $T=10^{-5}$ in an NVT ensemble using a Nos\'e-Hoover thermostat. 
We then ran NVT production runs. 
Their length was determined by the time needed for to decorrelate a term involving the local and global stress 
$\left\langle \sigma_{\alpha\beta}^{m}\sigma_{\gamma\delta}\right\rangle$, which was identified as a slowly decorrelating term and discussed 
by Mizuno \textit{et al.} \cite{Mizuno2013} and is defined in Section \ref{sec:elastic}. 
We found that there were no finite size effects, but, consistently with the observation made in Ref. [\cite{Mizuno2013}], much longer 
production runs are needed for larger systems. 
For a system of $N=48000$ particles, which is mainly used to perform the elastic modulus calculations in this study, 
the length of the production runs time was $\Delta t=3\times 10^5$, which corresponds to $1.5\times 10^7$ time steps.
The results shown in the paper are for the $N=48000$ particle system unless otherwise specified.
We observed very infrequent jumps in the energy and the pressure even at the very low temperature that we used,  $T=10^{-5}$. 
We attribute these jumps to transitions between the locally stable minima. In the analysis we only use 
a continuous portion of the trajectory that excludes the energy jumps. 

\subsection{Elastic Modulus Calculations}
\label{sec:elastic}

To measure the local elastic response, the system is equally partitioned into cells of size 
$w=3.30$, $4.54$, $6.05$, and $12.11$. 
Several methods have been proposed to define and calculate the local elastic constants. 
Here we use the so-called ``fully local'' approach described by Mizuno, Mossa, and Barrat \cite{Mizuno2013}. This approach was also used in other studies
 \cite{Mizuno2016,Mizuno2014,MizunoMossaBarratEPL2013}. 
For each box $m$ the volume averaged stress tensor is calculated as:
\begin{equation}
\sigma_{\alpha\beta}^m = -\rho^m T\delta_{\alpha\beta} + \frac{1}{w^3}\sum_{i<j} \frac{\partial U(r^{ij})}{\partial r^{ij}}\frac{r_{\alpha}^{ij} r_{\beta}^{ij}}{r^{ij}}\frac{q^{ij}_m}{r^{ij}}
\label{loc_stress_tensor}
\end{equation}
where, $\rho^m$ is the local number density of cell $m$, $T$ is the temperature, $\delta$ is the Kronecker delta and 
$r_{ij}=|\mathbf{r}_i-\mathbf{r}_j|$.  
The parameter $q^{ij}_m$ is the segment of the line joining $\mathbf{r}_i$ and $\mathbf{r}_j$ that lies within the box $m$. 
We use Greek subscripts to denote the Cartesian coordinates ($\alpha,\beta,\gamma,\delta=x,y,z$) and Roman superscripts to denote particle labels. 
The global stress tensor is given by:
\begin{equation}
\sigma_{\alpha\beta} = \frac{1}{V}\sum_{m} w^3 \sigma_{\alpha\beta}^m = -\hat{\rho}T\delta_{\alpha\beta} + \frac{1}{V}\sum_{i<j} \frac{\partial U(r^{ij})}{\partial r^{ij}}
\frac{r_{\alpha}^{ij} r_{\beta}^{ij}}{r^{ij}}.
\label{glob_stress_tensor}
\end{equation}
We  first calculate the local modulus $C_{\alpha\beta\gamma\delta}^{m}$  given by
\begin{eqnarray}
C_{\alpha\beta\gamma\delta}^{m} &=& C_{\alpha\beta\gamma\delta}^{Am} - C_{\alpha\beta\gamma\delta}^{Nm} \nonumber \\
&=& C_{\alpha\beta\gamma\delta}^{Bm} + C_{\alpha\beta\gamma\delta}^{Cm} + C_{\alpha\beta\gamma\delta}^{Km} - C_{\alpha\beta\gamma\delta}^{Nm} \nonumber \\
C_{\alpha\beta\gamma\delta}^{Bm} &=& \frac{1}{w^3}\left\langle\sum_{i<j}\left(\frac{\partial^2 U}{\partial r^{ij^{2}}} - \frac{1}{r^{ij}}\frac{\partial U}{\partial r^{ij}}\right)\frac{r^{ij}_{\alpha}r^{ij}_{\beta}r^{ij}_{\gamma}r^{ij}_{\delta}}{r^{ij^2}}\frac{q^{ij}_m}{r^{ij}}\right\rangle \nonumber \\
C_{\alpha\beta\gamma\delta}^{Cm} &=& -\frac{1}{2}\Big[2\left\langle \sigma_{\alpha\beta}^m\right\rangle \delta_{\gamma\delta} - \left\langle \sigma_{\alpha\gamma}^m\right\rangle \delta_{\beta\delta} \nonumber \\
&& - \left\langle \sigma_{\alpha\delta}^m\right\rangle \delta_{\beta\gamma} - \left\langle \sigma_{\beta\gamma}^m\right\rangle \delta_{\alpha\delta} - \left\langle \sigma_{\beta\delta}^m\right\rangle \delta_{\alpha\gamma}\Big] \nonumber \\
C_{\alpha\beta\gamma\delta}^{Km} &=& 2\left\langle\hat{\rho}^m\right\rangle T \left(\delta_{\alpha\gamma}\delta_{\beta\delta} + \delta_{\alpha\delta}\delta_{\beta\gamma}\right) \nonumber \\
C_{\alpha\beta\gamma\delta}^{Nm} &=& \frac{V}{T}\left(\left\langle \sigma_{\alpha\beta}^{m}\sigma_{\gamma\delta}\right\rangle - \left\langle \sigma_{\alpha\beta}^m\right\rangle \left\langle \sigma_{\gamma\delta}\right\rangle\right),
\label{tensor_eqns}
\end{eqnarray}
where $C_{\alpha\beta\gamma\delta}^{Am}$ is the affine contribution and $C_{\alpha\beta\gamma\delta}^{Nm}$ is the non-affine contribution.
While the non-affine contribution vanishes in perfect crystalline systems at zero temperature, it has a magnitude comparable to the affine  
term in amorphous systems\cite{Tanguy2002}. 
The brackets $\left\langle \cdots \right\rangle$ denotes an ensemble average. 
The Born contribution $C_{\alpha\beta\gamma\delta}^{Bm}$ to the affine term stems from the 
uniform displacement of all particles and it determines the instantaneous elastic modulus under such 
displacements\cite{Yoshimoto2004}. 
The $C_{\alpha\beta\gamma\delta}^{Cm}$ term is due to the initial stress having a finite value \cite{Mizuno2013}. 
The $C_{\alpha\beta\gamma\delta}^{Km}$ term is the kinetic energy contribution to the local elastic modulus tensor. Compared to the Born and the non-affine terms, the kinetic energy contribution to the elastic constant is negligible.

As described by Mizuno \textit{et al.} \cite{Mizuno2013}, the local bulk modulus $K^m$ is defined from the pressure-volume change and 
the five shear moduli $G^m_1$, $\cdots$, $G^m_5$, are defined from two pure shear and three simple shear deformations. These moduli are 
given by the following linear combinations of $C_{\alpha\beta\gamma\delta}^m$
\allowdisplaybreaks
\begin{eqnarray}
K^m &=& \left( C^m_{xxxx}+C^m_{yyyy}+C^m_{zzzz} +C^m_{xxyy} \right. \nonumber \\
&& \left. +C^m_{yyxx}+C^m_{xxzz}+C^m_{zzxx}+C^m_{yyzz}+C^m_{zzyy}\right)/9 \nonumber\\
G^m_1 &=& \left(C^m_{xxxx}+C^m_{yyyy}-C^m_{xxyy}-C^m_{yyxx}\right)/4 \nonumber \\
G^m_2 &=& \left[C^m_{xxxx}+C^m_{yyyy}+4C^m_{zzzz} +C^m_{xxyy}+C^m_{yyxx}\right. \nonumber\\
&& \left. -2\left(C^m_{xxzz}+C^m_{zzxx}+C^m_{yyzz}+C^m_{zzyy}\right)\right]/12 \nonumber\\
G^m_3 &=& C^m_{xyxy} \nonumber\\
G^m_4 &=& C^m_{xzxz} \nonumber\\
G^m_5 &=& C^m_{yzyz}.
\label{modulus_eqns}
\end{eqnarray}

The moduli are averaged over MD configurations that are separated by $t = 0.5$, \textit{i.e.} over $6\times 10^5$ time steps.

\section{Results}
Shear and bulk moduli describe the elastic response of the system to a small deformation. In simulations one can
determine these moduli through a deformation, or utilize the thermodynamic equations summarized in 
Eqns. (\ref{tensor_eqns}-\ref{modulus_eqns}) for the whole system, \textit{i.e.} 
when the system is only partitioned into one box. Here, we partition the system into several boxes and determine 
distributions of the moduli. We expect that the averages of these distributions should be equal to the values of the moduli obtained 
from deformation. To check this, we calculated the averages of the moduli for different 
box sizes $w$ and compared these results to the shear and bulk moduli obtained from deformation.   

\begin{figure}
\includegraphics[width=0.45\textwidth]{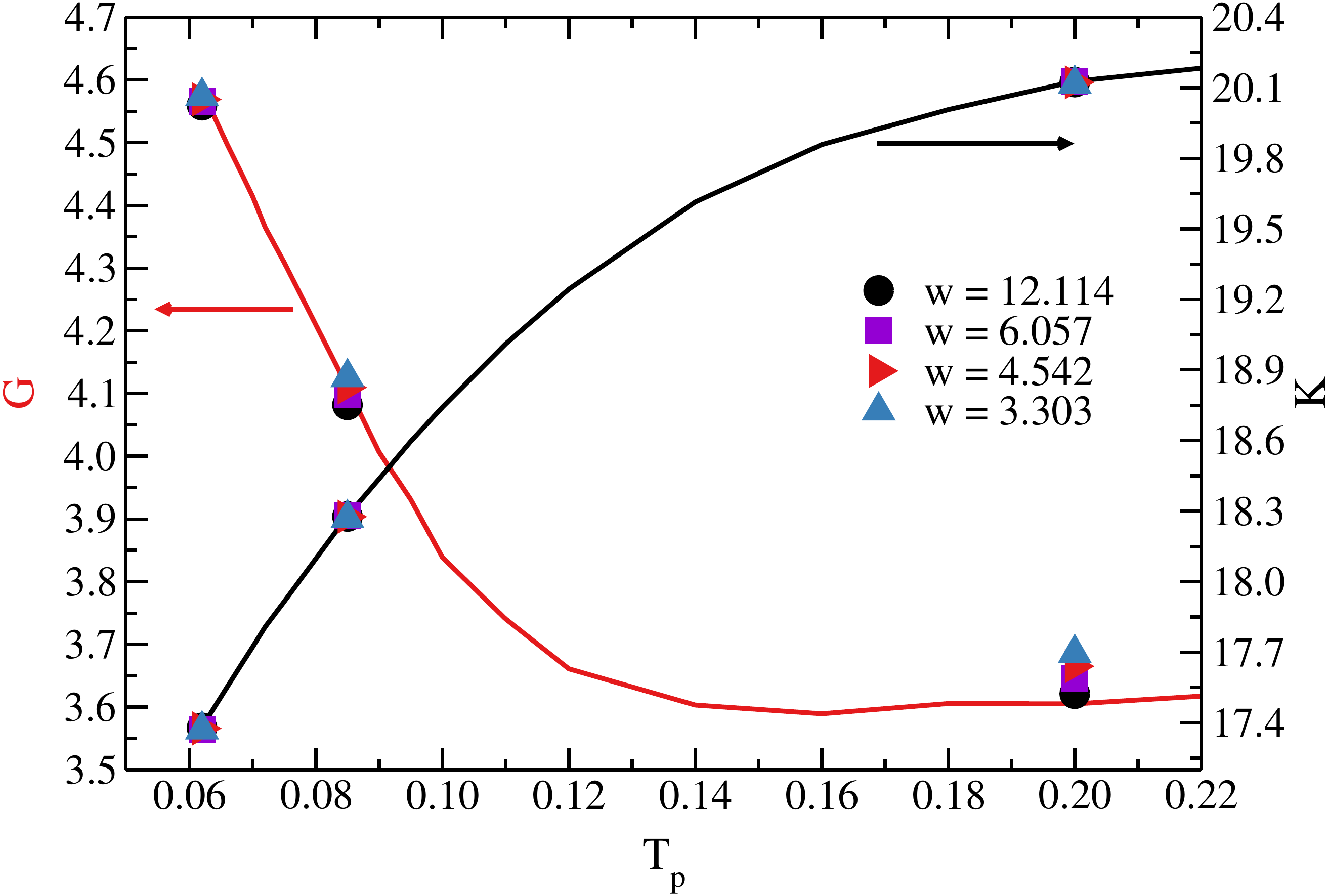}
\caption{\label{shear_bulk_avgs}Macroscopic shear (red line) and bulk (black line) moduli obtained by deforming the zero temperature (quenched) configurations 
as functions of the parent temperature. The symbols show the averages of the local shear and bulk moduli for different box sizes. 
The errorbars for the local moduli averages, not shown here, are smaller than or comparable to the size of the symbols.}
\end{figure}

Shown in Fig.~\ref{shear_bulk_avgs} are the shear modulus (left axis) and the bulk modulus (right axis) 
obtained from deforming the system (lines) and from the averages of the distributions of the local moduli (symbols) 
for different box sizes.
Up to the mode coupling temperature $T_{\text{MCT}}$ the global shear modulus $G$ 
changes very little with decreasing parent temperature $T_{\text{p}}$. Below $T_{\text{MCT}}$ it increases with decreasing $T_{\text{p}}$,
reaching a value approximately $27\%$ larger at the lowest parent temperature used.  
In contrast, the global bulk modulus $K$ monotonically decreases with decreasing $T_{\text{p}}$, reaching a value $7\%$ smaller at the lowest
parent temperature than at $T_{\text{MCT}}$.
The averages of the local shear $G^{\text{m}}$ and bulk $K^{\text{m}}$ moduli for different box sizes are very close to the moduli obtained from deformation. 
We do find, however, that at the largest parent temperature the averages of the
shear moduli are slightly larger than the value obtained from deformation, with the difference increasing systematically with decreasing 
box size. 

\begin{figure}
\includegraphics[width=0.45\textwidth]{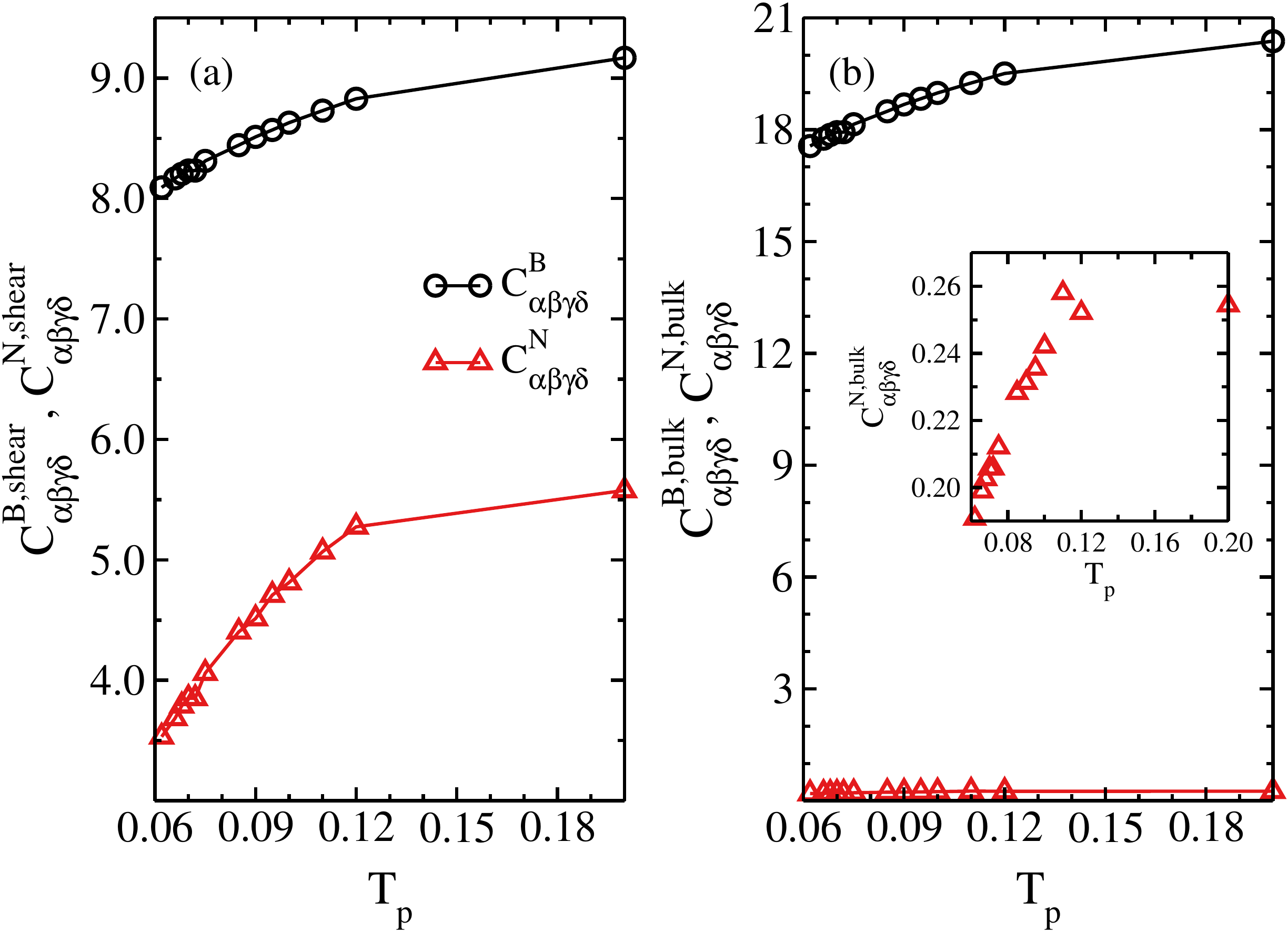}
\caption{\label{born_fluc_terms}The dependence of the Born and fluctuation terms on the parent temperature. Inset: rescaled data for the bulk fluctuation 
term. Both Born and fluctuation terms decrease with decreasing parent temperature, for both shear (a) and bulk (b) moduli. }
\end{figure}

We note that, as shown in Fig.  \ref{born_fluc_terms}, for both of the global shear and the global bulk moduli the Born and fluctuation 
terms in $C_{\alpha\beta\gamma\delta}$ decrease with decreasing $T_{\text{p}}$.
For the shear modulus, the fluctuation term decreases faster with decreasing $T_p$ than the Born term, and this  
leads to the increase in the shear modulus since the two terms are the same
order of magnitude. However, for the bulk modulus the fluctuation term
is  an order of magnitude smaller than the Born term, and thus a decrease in the Born term 
leads to a decrease of the bulk modulus.  
 
Although the average shear and bulk moduli are approximately independent of the box width $w$, one would expect to find some box width dependence of
the width of the moduli distributions. The dependence of the width of the distribution relative on 
the box size is an important parameter in the fluctuating elasticity theory. 
Mizuno \textit{et al.} found that the distributions of the individual shear moduli are almost identical and presented distributions averaged over the 
individual components. We found that the same fact is true for our system and also present distributions of the shear moduli averaged over the 
individual components.

\begin{figure}
\includegraphics[width=0.45\textwidth]{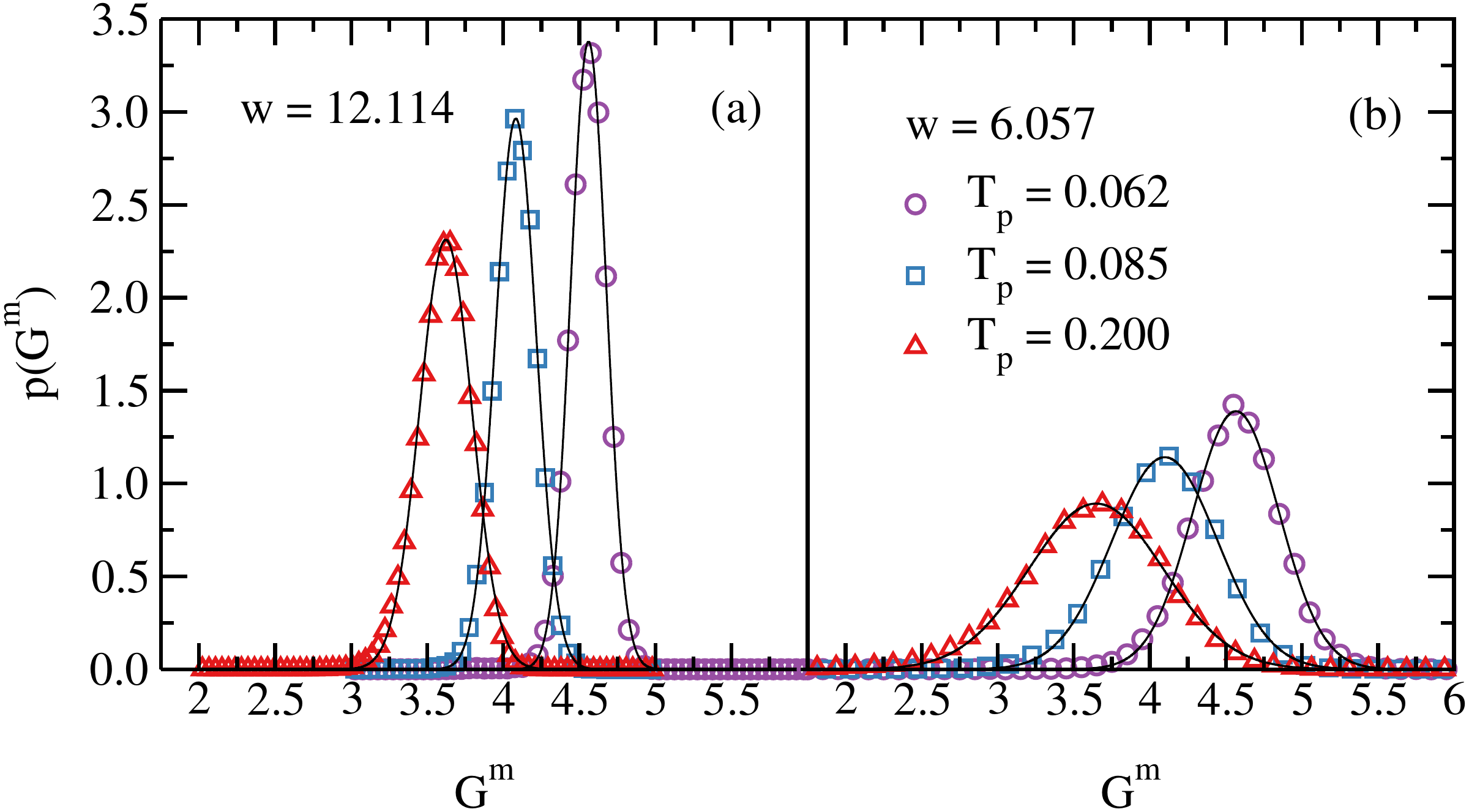}
\includegraphics[width=0.45\textwidth]{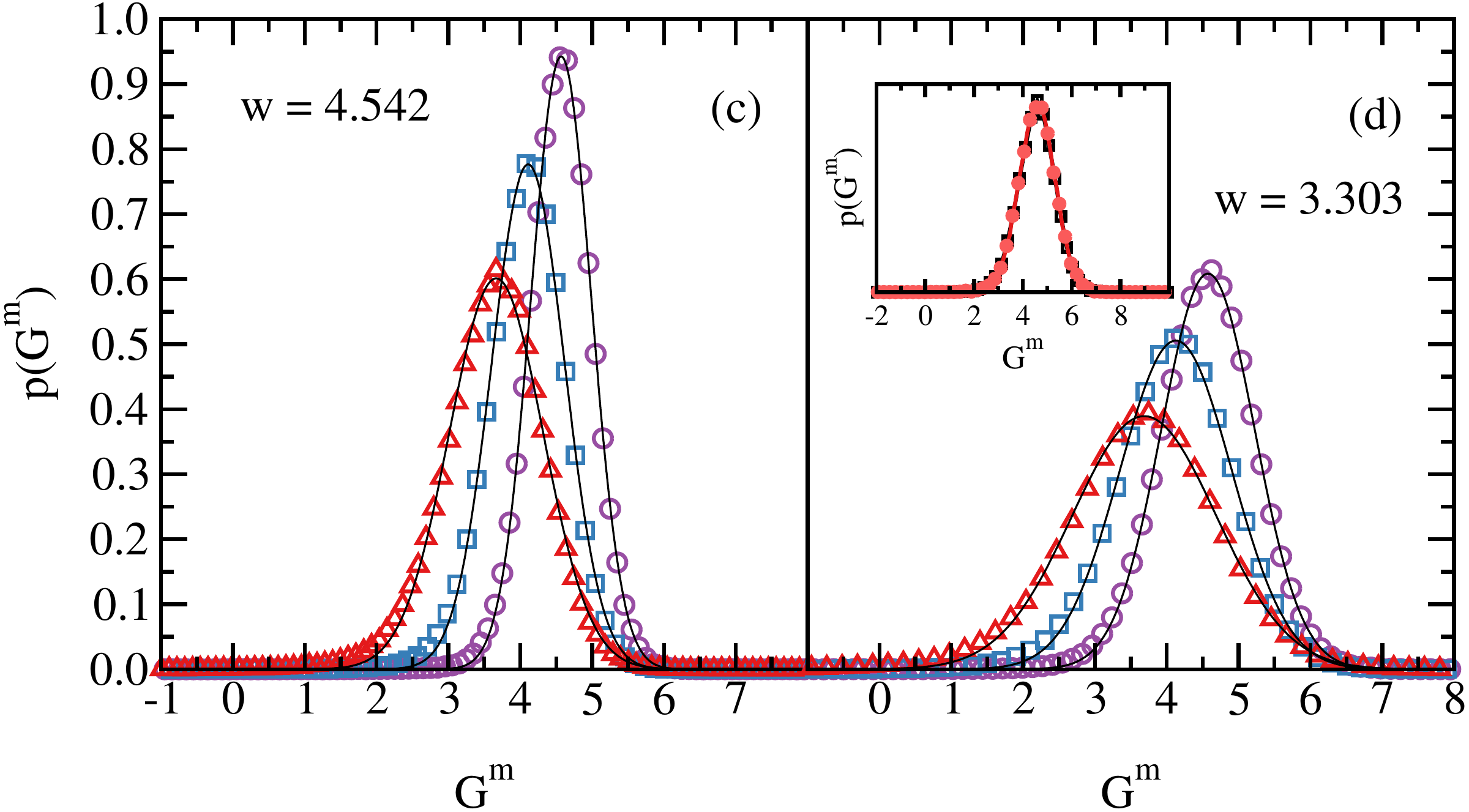}
  \caption{\label{shear_distributions}Distributions of local shear moduli for different box sizes: (a) $w=12.114$, (b) $w=6.057$, (c) $w=4.542$, (d) $w=3.303$. 
Each panel shows distributions for three different parent temperature, circles, $T_p=0.062$, squares, $T_p=0.085$ and triangles $T_p=0.200$. 
The solid lines show Gaussian  fits to the distributions.}
\end{figure}

Shown in Fig. \ref{shear_distributions} are probability distributions of the local shear modulus $G^{\text{m}}$ calculated for 
(a) $w$ = 12.114, (b) $w$ = 6.057, (c) $w$ = 4.543, and (d) $w$ = 3.303
for three parent temperatures $T_p$ = 0.062 (circles), 0.085 (squares), and 0.2 (triangles).
We note that there are no finite size effects, which we demonstrate in the inset to Fig.~\ref{shear_distributions}(d)
by calculating the distribution for $N=48000$ and $N=192000$ for a box of the same size.  
However, as discussed in Ref. [\cite{Mizuno2013}], 
the $\left\langle \sigma_{\alpha\beta}^{m}\sigma_{\gamma\delta}\right\rangle$ term 
converges very slowly for large systems. 
To characterize the width we fit the distributions to a Gaussian distribution, 
$A \exp\{ - 0.5 (G-G_0)^2/\sigma^2 \}$, 
where $G_0$ is the average shear modulus and $\sigma$ is the standard deviation.  
The fits are shown as straight lines in the figures. 
For all box sizes, including the smallest one with $w=3.303$ that only contains 
$\simeq 36$ particles, the shear moduli 
distributions are well described by Gaussian distributions. 

We can see two trends. First, with increasing stability the distribution becomes narrower. This is easily seen since
the peak of the distribution increases with decreasing width due to normalization of the distributions. Therefore, with increasing stability the
glass becomes more uniform, in the sense that the local shear moduli vary less between different boxes. The other
trend is that the width becomes broader with decreasing box size.
This result is intuitively expected. 

One noticeable property of some of these distributions is the appearance of regions with negative moduli. 
The regions with negative moduli are characterized as domains where the deforming force and the 
resulting response are in opposite directions\cite{Lakes2001}, which suggests that these
domains are unstable. However, with such small domains it is 
questionable if continuum elasticity is a valid description \cite{Tanguy2002}. 
Overall, at each box size the distributions with higher averages and smaller standard deviations 
(\textit{i.e.} the distributions of $T_{\text{p}}=0.062$) represent the more stable structure\cite{Fan2014}.

\begin{figure}
\includegraphics[width=0.45\textwidth]{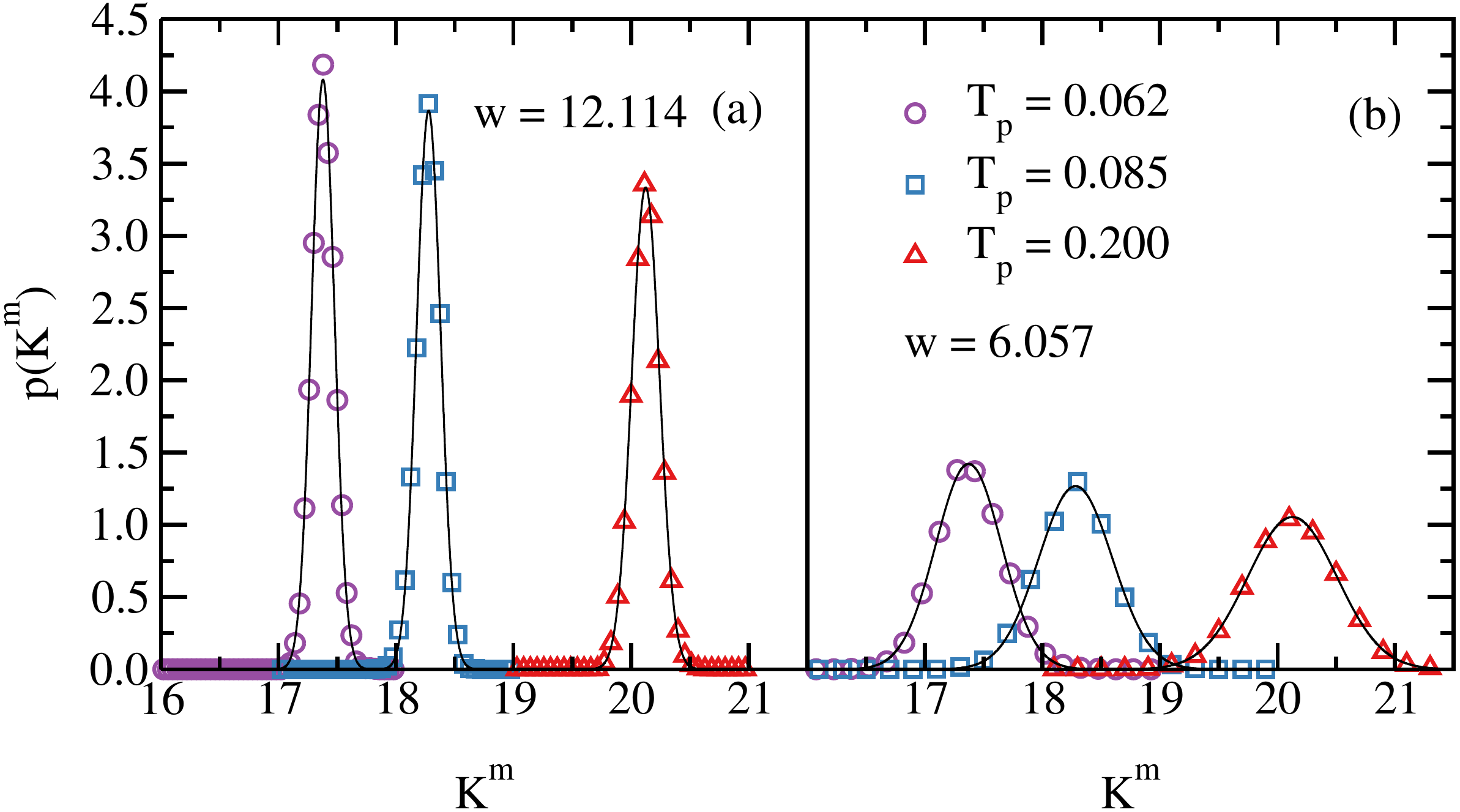}
\includegraphics[width=0.45\textwidth]{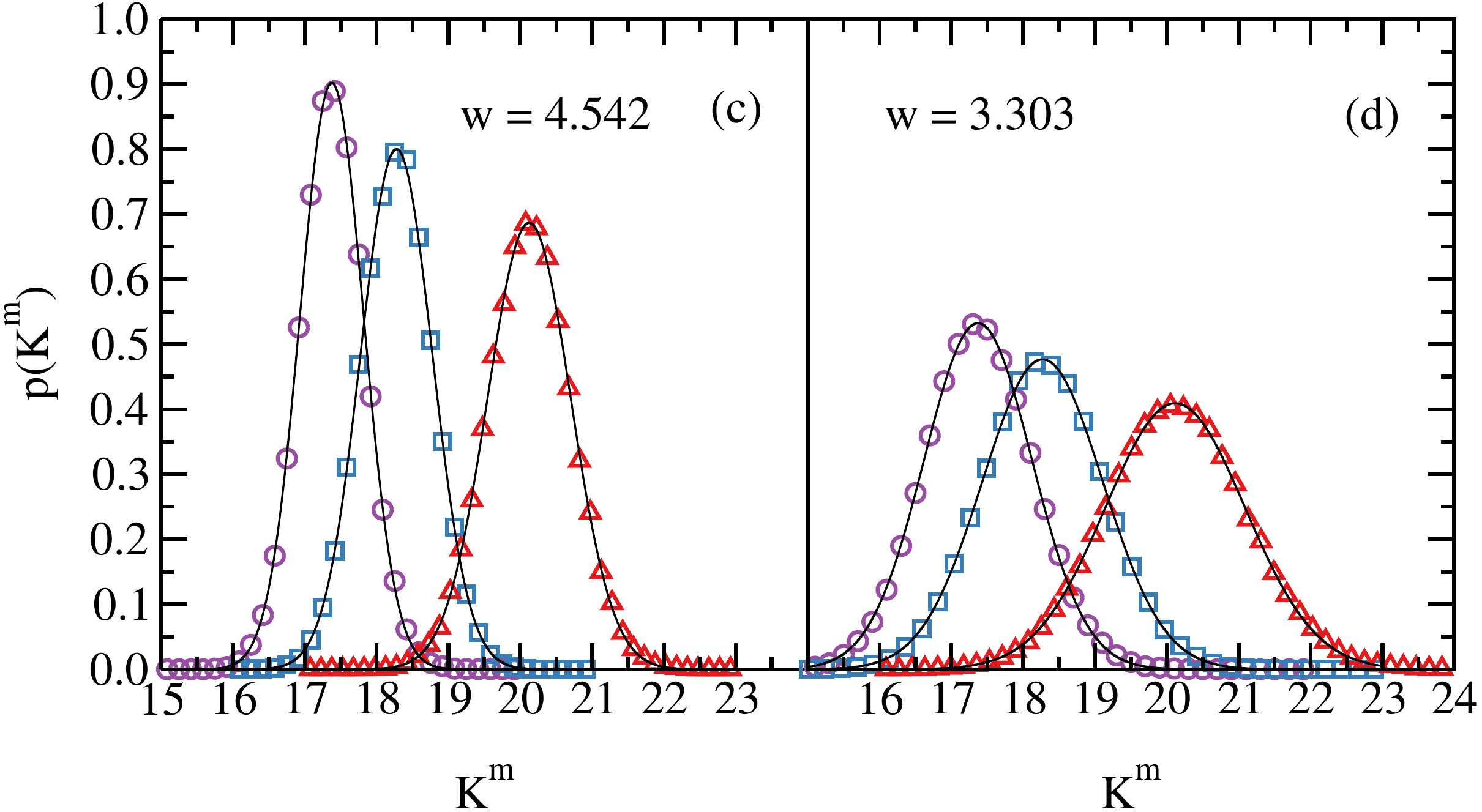}
\caption{\label{bulk_distributions}Distributions of local bulk moduli for different box sizes: (a) $w=12.114$, (b) $w=6.057$, (c) $w=4.542$, (d) $w=3.303$. 
Each panel shows distributions for three different parent temperature, circles, $T_p=0.062$, squares, $T_p=0.085$ and triangles $T_p=0.200$. 
The solid lines show Gaussian  fits to the distributions. }
\end{figure}

We also examined the distribution of the bulk modulus $K^{\text{m}}$, Fig. \ref{bulk_distributions}
for the same three parent temperature $T_{\text{p}}$ and box sizes $w$. We also  
find that the width of the distribution of $K^{\text{m}}$ decreases with decreasing parent temperature
and increases with decreasing box size. The lines in the figures are fits to a 
Gaussian distribution. Again, these results points to the bulk modulus becoming more uniform
with an increase of the stability. Since the bulk modulus is 3.5 to 5.5 times larger than the 
shear modulus (depending on stability), the change in the relative size of the distribution $\sigma_\Gamma/\Gamma$,
where $\Gamma = G$ or $K$ is much less for the bulk modulus. 

\begin{figure}
\includegraphics[width=0.45\textwidth]{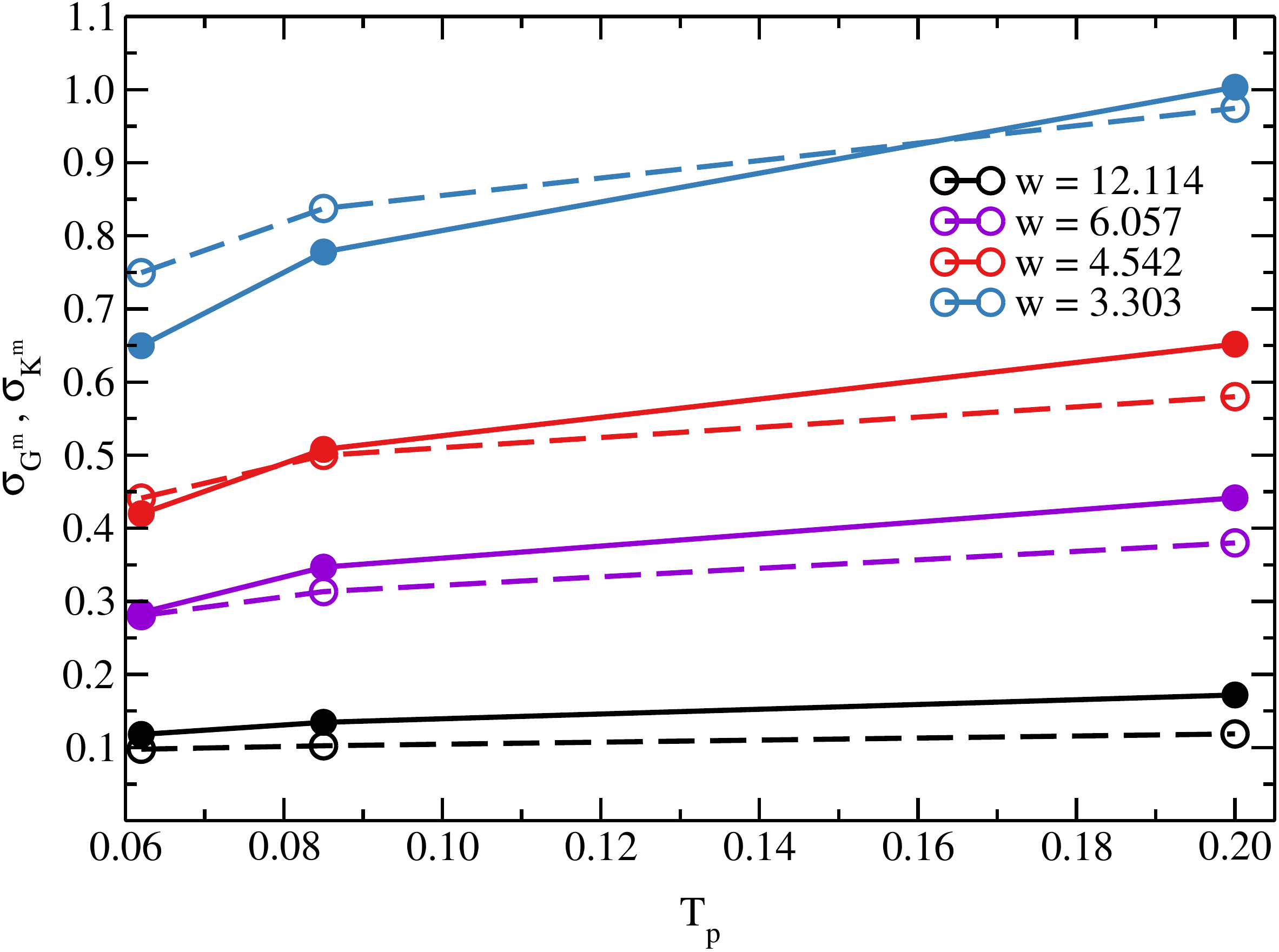}
\caption{\label{stdev_Tp}Dependence of the standard deviation of the local shear, $\sigma_{\text{G}^{\text{m}}}$, and bulk moduli, 
$\sigma_{\text{K}^{\text{m}}}$, on the 
parent temperature. The solid lines and filled symbols show $\sigma_{\text{G}^{\text{m}}}$ and the dashed lines and open symbols show $
\sigma_{\text{K}^{\text{m}}}$. The standard deviation $\sigma_{\text{G}^{\text{m}}}$ increases by 67\% for our smallest box size $w=3.303$ and 
50\% for our largest box size $w=12.114$. The standard deviation $\sigma_{\text{K}^{\text{m}}}$ increases by 33\% for our smallest box size
and 7.1\% for our largest box size. Since $K > G$, this signifies a much larger relative change in $\sigma_{\text{G}^{\text{m}}}$ than 
$\sigma_{\text{K}^{\text{m}}}$.}
\end{figure}

We summarize the parent temperature and box size dependence of the standard deviation of the distributions of the local moduli 
in Fig.~\ref{stdev_Tp}. 
The closed symbols are the results for the shear moduli and the open symbols are results for the bulk modulus. 
The increase in $\sigma_{G^m}$ upon decreasing the box of size from $w$ = 12.114 to $w$ = 3.303 is a factor of 5.5 
for $T_p$ = 0.2 and 5.8 for $T_p$ = 0.062.  Similarly, the decreases of $\sigma_{G^m}$ with 
parent temperature for a fixed box size is 31\% for $w$ = 12.114 and 35\% for $w$ = 3.303.
 
We calculated the disorder parameter of fluctuating elasticity theory \cite{Schirmacher2006,Schirmacher2007,Marruzzo2013},
$\gamma_G = \rho w^3 \sigma_{G^m}^2/\left<G^m\right>^2$ for the different box sizes. We found that the disorder parameter 
varies with box size. For our most stable glass, $T_p = 0.062$, $\gamma_G = 1.24$ for $w=12.1$ and $\gamma_G = 0.90$ for $w=3.3$.
The disorder parameters differ by approximately $38\%$. This box size dependence of the disorder parameter makes it unclear
if this is a proper parameter to be used as input to a theory of excitations in glasses. 

The disorder parameter does increase dramatically
with decreasing stability for a fixed box size. The disorder parameter increases by a factor of 3.4-3.9, depending on box size, 
when we compare our most stable glass, $T_p = 0.062$, to our least stable glass, $T_p = 0.2$.
For our least stable glass, disorder parameters are of similar magnitude as thouse found 
by Mizuno, Ruocco, and Mossa \cite{Mizuno2019} in their $T=0$ glass. 
    
We note that the change in the variation of the local elastic moduli, \textit{i.e.} of the heterogeneity of the local elasticity, with
the changing stability found in this work
is much larger than that estimated by Pogna \textit{et al.}\ for hyperaged amber. In the latter study a decrease of only 5\%
was estimated upon a very large increase in the stability. We note that the change in the 
variation of the elastic constants reported by Pogna \textit{et al.}\ was obtained indirectly, by fitting measured vibrational densities of
states to the predictions of the fluctuating elasticity theory. Thus, the accuracy of their inferred change of the variation of the local
elastic moduli depends on 
accuracy of the fluctuating elasticity model that they used. We find that there is probably a stonger dependence of the variation of the 
elastic constants on the glass' stability than that inferred from fluctuating elasticity theory.

To characterize the spatial correlations of local shear moduli, which also enter into the fluctuating elasticity theory \cite{Schirmacher2010},
we calculated the correlation function
\begin{equation}
g_{GG}(r) = \sum_m \sum_{n} \left(\left<G^m G^n\right> -\left<G^m\right>\left<G^n\right>\right)\delta(r - |\mathbf{r}_m - \mathbf{r}_n|),
\end{equation}  
where $\mathbf{r}_n$ is the coordinate for the center of a box used to calculate the elastic moduli. We
used 3000 particle systems to calculate $g_{GG}(r)$ and checked that the calculation was consistent with 
results for 48000 particle systems. It is important 
to recognize the fact that the boxes used in this calculation may overlap (in order to get results for distances $r$ smaller than the box size), 
and thus boxes may share some of the same particles and their elastic moduli are necessarily correlated. Therefore, there
are trivial correlations in $g_{GG}(r)$ due to overlapping boxes. We show $g_{GG}(r)$ for 
our most stable glass, $T_p = 0.062$, for four different box sizes $w$.  We find that only the trivial correlations 
exists and $g_{GG}(r)$ decays to near zero at the size of the box, which is indicated by the vertical lines in the figure. 

\begin{figure}
\includegraphics[width=0.45\textwidth]{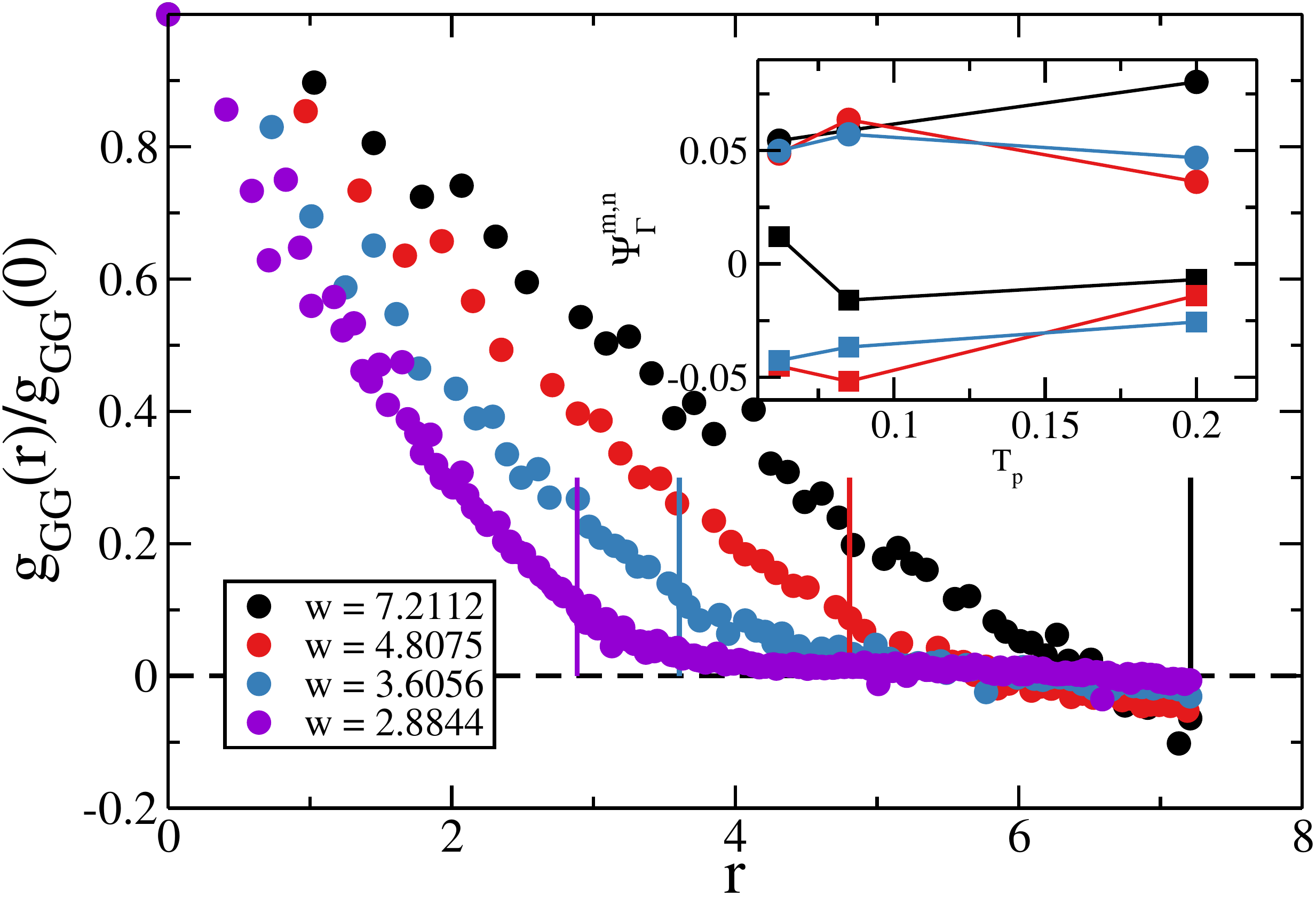}
\caption{\label{moduli_corr}The spatial correlations of the shear modulus $G$ for 
a 3000 particle system and for our most stable glass, 
$T_p = 0.062$. The different colors denote different 
box sizes $w=7.2112$ (black), 4.8075 (red), 3.6056 (blue), and 2.8844 (purple). The vertical lines 
indicate the box sizes. At these points the trivial correlations disappear. The inset shows
the correlation parameter $\Psi_G^{m,n}$ (circles) and $\Psi_K^{m,n}$ (squares) for the box sizes $w=6.075$ (black),
4.542 (red), and 3.028 (blue) as a function of parent temperature ($N=48000$). The correlation parameter is small and there is no clear 
box size or parent temperature dependence.}
\end{figure}

To explore further if there are spatial correlations for the shear modulus and the bulk modulus at every temperature and every
box size, we calculate the cross correlations of neighboring non-overlapping boxes. 
To this end we calculate the correlation parameter
\begin{equation}
\Psi_\Gamma^{m,n}=\left|\left\langle\left(\frac{\Gamma^m - \Gamma}{\sigma_{\Gamma^m}}\right)
\left(\frac{\Gamma^n-\Gamma}{\sigma_{\Gamma^n}}\right)\right\rangle_m\right|
\label{corr_parameter}
\end{equation}
where, $\left\langle\cdots\right\rangle_m$ denotes an average over all the boxes and box $n$ is one of the
six nearest neighbors of
box $m$ and $\Gamma = G$ or $K$. 
A correlation parameter close to $0$ indicates no significant correlation and a value of 
$1$ indicates perfect correlation. 
In the inset to Fig. \ref{moduli_corr} we show $\Psi_G$ (circles) and $\Psi_K$ (squares) for box sizes of $w=6.075$ (black), 4.542 (red), and 3.028 (blue).
The values of $\Psi_\Gamma$ are all close to zero and there are no noticeable trends with box size or parent temperature.
This leads us to conclude that the elastic moduli, calculated using this fully local approach, do not exhibit any spatial correlations. 
We also examined correlations of $G_n^m$ where $n=1 ... 5$ found in Equation \ref{modulus_eqns} and found the same trends, 
i.e.\ only trivial correlations. We note that there are other methods to calculate local elastic moduli \cite{Mizuno2013}, and
these other methods may indicate that the moduli are spatially correlated. 

This conclusion is at odds with the result of Gelin \textit{et al.}\ \cite{Gelin2016} who reported that
the elastic correlations decayed as $r^{-2}$ for a two dimensional glass-forming system different from the system used here. 
We note that Gelin \textit{et al.} used a different way to define local elastic moduli.  
However, Mizuno and Ikeda \cite{MizunoIkedaPhysRevE2018} utilized the same method as Gelin \textit{et al.}
for yet another, different two dimensional system and found that the stress correlations decay as $r^{-2}$, but
the elastic moduli correlations does not show the same long range correlations.     

\section{Conclusions}

We examined the structural heterogeneities, including local and global elastic moduli, of glassy systems prepared from parent systems at different initial 
temperatures. Our calculations showed that the glass has a rather mild 27\% increase of the local shear modulus,  and a smaller 7\% decrease on local bulk modulus
compared to their values at the mode-coupling temperature with decreasing parent temperature. More importantly, we found that 
the local shear and the local bulk moduli become more uniform with decreasing parent temperature and thus stability of the glass. 
This finding is consistent with the recent report on the stability and sound attenuation of stable glasses\cite{WangBerthierFlennerGuanSzamelSoftMatter2019}. 
Sound attenuation increases with an increase in the fluctuations of the local elasticity, and hence with a decrease of the stability. 
Our results are in qualitative agreement with fluctuating elasticity 
theory \cite{Schirmacher2006,Schirmacher2007,Marruzzo2013}, which predicts an increase of sound attenuation and the observed Rayleigh-like 
$k^4$ scaling for small wavevectors \cite{WangBerthierFlennerGuanSzamelSoftMatter2019,MizunoIkedaPhysRevE2018}.  

Our results are also qualitatively consistent with recent experimental work by Pogna \textit{et al.}\ on hyperaged amber\cite{Pogna2019}, 
which showed that the elastic matrix becomes more homogeneous with increased stability,
corresponding to a 
smaller $T_{\text{p}}$ and a narrower moduli distribution in our study.  However, we find
that the local moduli are not spatially correlated. Pogna \textit{et al.}\ inferred a 22\% increase in the length scale
characterizing elastic correlations. The same work reported on an increase of the elastic 
moduli fluctuation length scale in the more stable amorphous medium. This result, however, remains at variance with the findings of our study,
where there is no discernible length scale associated with elasticity and there is no long range decay of elastic correlations. The lack of long range
decay is also at odds with the study of Gelin \textit{et al.}\cite{Gelin2016}, but agrees with the conclusions of Mizuno and Ikeda \cite{MizunoIkedaPhysRevE2018}. 


Our results suggest that the current version of fluctuating elasticity theory is not a quantitatively accurate description of sound attenuation and the boson peak 
in amorphous solids, even though it makes qualitatively accurate predictions. A similar conclusion was drawn by Caroli and Lama\^itre \cite{Caroli2019}, 
who developed a full tensorial fluctuating elasticity theory and found that it underestimates the sound attenuation by about two orders of magnitude.
Further theoretical work is warranted to properly describe the interplay of sound attenuation and elastic heterogeneities.   

\providecommand*{\mcitethebibliography}{\thebibliography}
\csname @ifundefined\endcsname{endmcitethebibliography}
{\let\endmcitethebibliography\endthebibliography}{}

\end{document}